\documentstyle[12pt]{article}

\newcommand{\be}{\begin{equation}}
\newcommand{\ee}{\end{equation}}
\newcommand{\bn}{\begin{eqnarray}}
\newcommand{\en}{\end{eqnarray}}
\newcommand{\bd}{\begin{displaymath}}
\newcommand{\ed}{\end{displaymath}}

\begin{document}

\begin{flushright}
QMW/95-5
\\hep-th/9502016
\end{flushright}
\vspace{0.5cm}
\begin{center}
{\Large \bf Massive quantum fields in a conical
  background}\footnote{Work supported by a grant from the
  CNPq/Brasil.}
\\
\vspace{1cm}
{\large E.S. Moreira Jnr.}\footnote{e-mail: e.moreira@qmw.ac.uk}
\\
\vspace{0.3cm}
{\em Physics Department,} \\
{\em Queen Mary and Westfield College,}   \\
{\em Mile End Rd., London E1 4NS, UK}  \\
\vspace{0.3cm}
{\large January, 1995}
\end{center}
\vspace{1cm}

\begin{abstract}
Representations of the
Klein-Gordon and Dirac propagators are determined
in a $N$ dimensional conical background for massive fields
twisted by an arbitrary angle $2\pi\sigma$.
The Dirac propagator is shown to be obtained
from the Klein-Gordon propagator twisted by
angles $2\pi\sigma\pm {\cal D}/2$ where ${\cal D}$ is
the cone deficit angle.
Vacuum expectation values are determined
by a point-splitting method
in the proper time
representation of the propagators.
Analogies with the Aharonov-Bohm effect
are pointed out throughout the paper and
a conjecture on an extension to fields
of arbitrary spin is given.
\end{abstract}

\section{Introduction}
Recently interest in quantum fields
in a conical spacetime background
has been renewed with the
possibility of application to the evaluation
of geometric entropy \cite{dow94}.
The subject has been studied
since the late seventies \cite{dow77},
and investigations became rather intense in
the mid-eighties
[3-8]
due to the importance that cosmic
strings may have in cosmology \cite{vil85}.
The literature is mainly concerned with a massless
scalar field.

In studying particles and fields on a cone,
one meets features which resemble
in a great deal those in the
Aharonov-Bohm (A-B) set up \cite{aha59}.
The roots of this analogy are in the
gauge theory aspects of gravity which
are very special in 3 dimensions, as shown
in \cite{ger90} and references therein.
Field strengths are
concentrated on the symmetry axis,
vanishing everywhere else.
However, the non vanishing affine connections
tell the rest of the spacetime that there is
a singularity at the symmetry axis.
That is the source of
the non trivial A-B effect.

By performing a convenient gauge transformation,
one can replace the affine connections almost
everywhere by an unusual boundary condition \cite{ger90}.
In the A-B situation for example the
vector potential can be gauged away
everywhere except on some ray.
In the new gauge the fields are ``twisted''
by an angle
$2\pi\sigma$,
\bd
{\cal F}(\theta+2\pi)=
\exp\{i2\pi\sigma\}{\cal F}(\theta),
\ed
where
$\theta$
is the polar angle and
$\sigma$
is the flux parameter \cite{dow77,fel81}.

This paper is concerned with massive scalar and
spinor fields in a $N$ dimensional conical
spacetime. The fields are taken to be ``twisted''
by an angle $2\pi\sigma$. In section 2 the geometry
of a conical spacetime generated by a source
carrying spin (spinning cone) is described. The proper time
representation of the Klein-Gordon propagator is
obtained in section 3. Evaluating the integration over the
proper time, another representation is presented.
The rest of the paper refers to a spinless cone.

In section 4 the Dirac propagator is expressed in
terms of the Klein-Gordon propagator.
Its expression is derived using a $N$-bein for
which the spin connection vanishes almost
everywhere. This and other resemblances
with the A-B effect are also discussed.

Section 5 is concerned with vacuum expectation
values. The approach used is the one in \cite{can79},
which has been used earlier to work out the
energy momentum tensor
$\langle T^{\mu}{}_{\nu}(x)\rangle$
of a conformal scalar field
in a 4 dimensional conical spacetime
\cite{hel86,mat90a}. All the quantities
can be obtained from the twisted Klein-Gordon
propagator.

General expressions showing dependence on
the radial coordinate $\rho$ are given
for the vacuum fluctuations
$\langle\phi ^{2}(x)\rangle$ and
$\langle T^{\mu}{}_{\nu}(x)\rangle$
of a massive ($M$) scalar
field coupled with the curvature scalar, when
$M\rho<<1$. These quantities are evaluated
in four dimensions ($N=4$) showing the dependence on the
deficit angle ${\cal D}$ and on the
twist parameter $\sigma$. A mass correction
in
$\langle\phi ^{2}(x)\rangle$
is determined for a untwisted
$\phi(x)$. For spinors, the energy density
$\langle T_{0}{}_{0}(x)\rangle$
is given in $N=4$ when
$M\rho<<1$.

The paper ends with some
possible extensions of the work.
In particular a conjecture on
the Feynman propagator of a field of
arbitrary spin is given.
Throughout the paper
$c=\hbar=1$
and $G=1/4$.

\section{The background}
A $N$ dimensional spinning cone
\cite{des84,ger89a}
is characterized by the Minkowski
line element written in cylindrical coordinates
\be
ds^{2}=d\bar{t}^{2}-d\rho^{2}-\rho^{2}d\varphi^{2}-d{\bf z}^{2},
\label{le}
\ee
and by the identification
\be
\left(\bar{t},\rho ,\varphi,  {\bf z}\right)
\sim  \left(\bar{t}+2\pi S,\rho ,\varphi +2\pi\alpha ,{\bf z}\right),
\label{id}
\ee
where
${\bf z}:= (z_{1},...,z_{N-3})$,
$S\geq 0$ is the spin density of the source on the
cone symmetry axis ($\rho=0$),
and $0<\alpha\leq 1$ is the cone parameter.
The latter is related to the deficit angle by
${\cal D}= 2\pi \left(1-\alpha \right)$ which
is proportional to the
mass density of the source,
$\mu={\cal D}/2\pi$.
The unusual identification (\ref{id}) encapsulates
the fact that (\ref{le}) hides a singularity
on the symmetry axis. Clearly when $\alpha=1$ and
$S=0$ the spacetime
becomes the Minkowski one.

\section{The Klein-Gordon propagator}
The Feynman propagator for a spin 0 field with
mass $M$ in the
background described above is solution of \cite{dav82}
\be
\left(\Box_{x}+M^{2}\right)G_{{\cal F}}(x,x')=
-\frac{1}{\rho}\delta\left(x-x'\right),
\label{fe0}
\ee
where
\bd
\Box_{x} =\frac{\partial^{2}}{\partial \bar{t}^{2}}
-\frac{1}{\rho}\frac{\partial}{\partial
\rho}\left(\rho\frac{\partial}{\partial \rho}\right)
+\frac{ {\bf L}^{2}}{\rho^{2}}
-{\bf \nabla}_{{\bf z}}
\ed
with
${\bf L}:=-i \partial/\partial \varphi$
and
${\bf \nabla}_{{\bf z}}:=
\sum_{i=1}^{N-3}\partial ^{2}/\partial z_{i}^{2}$.
Since the spacetime is flat around the cone symmetry
axis, the coupling with the curvature scalar does not
contribute in (\ref{fe0}).
If the field is ``twisted'' by an angle $2\pi\sigma$,
it follows from (\ref{id}) that
\be
G_{{\cal F}}(\bar{t}+2\pi S,\varphi+2\pi\alpha)=
\exp\{i2\pi\sigma\}G_{{\cal F}}(\bar{t},\varphi),
\label{bc1}
\ee
where the other coordinates have been
omitted.
The twisted boundary condition
(\ref{bc1}) implies that the propagator must
vanish on the cone symmetry axis
for non-integer values of the twist parameter $\sigma$,
otherwise one gets an inconsistency by shrinking
a loop around $\rho=0$ \cite{dir36}
\footnote{In fact this is the case only when the
propagator is finite at $\rho=0$.
There are other choices of boundary conditions
which diverge at $\rho =0$ \cite{kay91}.
They will not be considered here.}.

If $0 \leq\varphi < 2\pi\alpha$
the $\delta$ function in (\ref{fe0}) is
\bd
\delta(x-x')= \delta\left(\bar{t}-\bar{t}'-S(\varphi -\varphi ')/\alpha
\right)\delta\left(\rho -\rho '\right)
\delta(\varphi-\varphi ')
\delta\left({\bf z}-{\bf z}'\right),
\ed
which is adequate to the boundary condition (\ref{bc1}).
Since $\Box _{x}$ is just the D'Alembertian in Minkowski
spacetime written in cylindrical coordinates, the non-trivial
geometry manifests itself only through (\ref{bc1})
\footnote{When $\sigma$ is identified with
a flux parameter, the field is taken to be charged
and the interaction with the localized
magnetic flux is also expressed
only through (\ref{bc1}).}.

In order to get the proper time representation of $G_{{\cal F}}(x,x')$
one needs the eigenfunctions of the operator
$\Box_{x}+M^{2}$ satisfying (\ref{bc1}). They are given by
\be
\psi_{\omega,\kappa,m,{\bf k}}(x)=
\frac{1}{(2\pi)^{(N-1)/2}\alpha ^{1/2}}
J_{|m+\sigma +\omega S|/\alpha}(\kappa\rho)
e^{i[{\bf k}\cdot {\bf z}- \omega\bar{t}
+(m+\sigma +\omega S)\varphi/\alpha]},
\label{eig1}
\ee
where $m$ is a integer, $\omega$ and ${\bf k}$ are real numbers,
$\kappa$ is a positive real number
and $J_{\nu}$ denotes a Bessel function of the first kind.
The corresponding eigenvalues are
$E_{\omega ,\kappa ,{\bf k}}=
\kappa ^{2}+{\bf k}^{2}-\omega ^{2}+M^{2}$.
$\omega S$ and $\sigma$
play the same role in (\ref{eig1}), namely to
shift the eigenvalues $m/\alpha$ of ${\bf L}$.
This analogy has also been pointed out in
\cite{ger89a,ger90} in related contexts.

Using the completeness relation of the Bessel functions
\bd
\int_{0}^{\infty}dk\ kJ_{\nu}(k\rho)J_{\nu}(k\rho')=
\frac{1}{\rho}\delta(\rho-\rho')
\ed
and the Fourier expansion (Poisson's formula)
\bd
\sum_{m=-\infty}^{\infty}\delta(\theta +2\pi m)=
\frac{1}{2\pi}\sum_{m=-\infty}^{\infty}\exp\left\{im\theta\right\},
\ed
one obtains
\be
\sum_{m=-\infty}^{\infty}
\int_{-\infty}^{\infty}d\omega\
\int_{-\infty}^{\infty}d{\bf k}
\int_{0}^{\infty}d\kappa\ \kappa
\psi_{\omega,\kappa,m,{\bf k}}(x)
\psi^{\ast}_{\omega,\kappa,m,{\bf k}}(x')=
\frac{1}{\rho}\delta\left(x-x'\right),
\label{cr1}
\ee
which is the completeness relation of the
eigenfuntions
$\psi_{\omega,\kappa,m,{\bf k}}(x) $.
Considering (\ref{cr1}) and by direct
application of $\Box_{x} + M^{2}$,
it can be shown that
$G_{{\cal F}}(x,x')$ is given by
\bn
G_{{\cal F}}^{(N,S,\alpha,\sigma)}(x,x')&=&
-i\int_{0}^{\infty}dT\
\sum_{m=-\infty}^{\infty}
\int_{-\infty}^{\infty}d\omega\
\int_{-\infty}^{\infty}d{\bf k}
\nonumber\\
&&\times\int_{0}^{\infty}d\kappa\ \kappa
e^{-iTE_{\omega ,\kappa ,{\bf k}}}
\psi_{\omega,\kappa,m,{\bf k}}(x)
\psi^{\ast}_{\omega,\kappa,m,{\bf k}}(x'),
\label{pro1}
\en
where $E_{\omega ,\kappa ,{\bf k}}$
is taken to have an infinitesimal
negative imaginary part to make the integration
over $T$ converge.
$G_{{\cal F}}^{(4,0,\alpha,0)}(x,x')$,
$G_{{\cal F}}^{(4,0,\alpha,1/2)}(x,x')$
and
$G_{{\cal F}}^{(4,S,\alpha,0)}(x,x')$
reproduce the expressions in
\cite{hel86}, \cite{mat90a} and \cite{mat90b}
respectively.
Evaluating the integrations over ${\bf k}$
and $\kappa$ \cite{grad80}, (\ref{pro1}) yields
\bn
G_{{\cal F}}^{(N,S,\alpha,\sigma)}(x,x')&=&
\int_{0}^{\infty}dT\ \frac{\left(T/i\pi\right)^{1/2}}
{\alpha(4\pi iT)^{N/2}}
e^{-i\left\{[-(\rho^{2}+\rho'^{2})
-({\bf z}-{\bf z}')^{2}]/4T + M^{2}T\right\}}
\nonumber\\
&&\times\int_{-\infty}^{\infty}d\omega\
e^{iT\omega ^{2}-i\omega(t-t')}
\sum_{m=-\infty}^{\infty}I_{|m+\sigma +\omega S|/\alpha}
\left(\rho\rho'/2iT\right)
e^{i(m+\sigma)\left(\varphi-\varphi'\right)/\alpha},
\nonumber
\en
where
$I_{\nu}$ denotes a modified Bessel function
of the first kind, the time coordinate has been redefined
\footnote{In terms of the new time, (\ref{le}) becomes
$ds^{2}=(dt+Sd\varphi/\alpha)^{2}-d\rho^{2}
-\rho^{2}d\varphi^{2}-d{\bf z}^{2}$ and (\ref{id}) becomes
$\left(t,\rho ,\varphi,  {\bf z}\right)
\sim  \left(t,\rho ,\varphi +2\pi\alpha ,{\bf z}\right)$.}
as $t:=\bar{t}-S\varphi /\alpha$, and the fact that
the propagator transforms like a scalar at
$x$ and $x'$ (biscalar) has been considered.
When the cone is spinless the integration over
$\omega$ can be performed,
\bn
G_{{\cal F}}^{(N,0,\alpha,\sigma)}(x,x')&=&
\int_{0}^{\infty}\frac{dT}
{\alpha(4\pi iT)^{N/2}}
e^{-i\left\{[(t-t')^{2}-(\rho^{2}+\rho'^{2})
-({\bf z}-{\bf z}')^{2}]/4T + M^{2}T\right\}}
\nonumber\\
\label{pro3}
&&\times\sum_{m=-\infty}^{\infty}I_{|m+\sigma |/\alpha}
\left(\rho\rho'/2iT\right)
e^{i(m+\sigma)\left(\varphi-\varphi'\right)/\alpha}.
\en

Recalling the Fourier expansion of a plane wave
\bd
\exp\left\{z\cos\theta\right\}=
\sum_{m=-\infty}^{\infty}I_{|m|}(z)e^{im\theta},
\ed
one sees that
$G_{{\cal F}}^{(N,0,1,0)}(x,x')$
reduces to the proper time expression
for the Feynman propagator in the Minkowski
spacetime
\footnote{This representation resembles
a first quantization propagator integrated over
the ``time'' $T$.},
\be
G_{{\cal F}}^{(N,0,1,0)}(x,x')=
\int_{0}^{\infty}\frac{dT}
{(4\pi iT)^{N/2}}
e^{-i\left[(X-X')^{2}/4T + M^{2}T\right]},
\label{pro4}
\ee
with $X^{\mu}=(t,\,x=\rho\cos\varphi,\,y=\rho \sin \varphi,
\,{\bf z})$ being genuine Minkowski coordinates
\footnote{When $\alpha\neq 1$ $X^{\mu}$ are
singular ``Minkowski''coordinates. In terms
of these coordinates the metric tensor is Minkowski
everywhere except on $\varphi =0\sim 2\pi\alpha$
where $X^{\mu}$ are discontinuous. This is analogous
to what happens in the A-B set up.}.
The usual representation is obtained from
(\ref{pro4}) noticing that
$G_{{\cal F}}^{(N,0,1,0)}(x,x')=
\int_{0}^{\infty}(dT/i(2\pi)^{N})
\linebreak[4]
\int_{-\infty}^{\infty}d^{N}K
\exp\{-i\left[K(X-X') -(K^{2}- M^{2})T\right]\}$.
Performing the integration over $T$,
\be
G_{{\cal F}}^{(N,0,1,0)}(x,x')=
\int_{-\infty}^{\infty}
\frac{d^{N}K}{(2\pi)^{N}}
\frac{e^{-iK(X-X')}}{K^{2}-M^{2}+i\epsilon}.
\label{ur}
\ee

Another representation of
$G_{{\cal F}}^{(N,S,\alpha,\sigma)}(x,x')$
can be obtained from
(\ref{pro1}) by performing the integration
over the ``proper time'' $T$ \cite{grad80}
and over {\bf k},
\bn
G_{{\cal F}}^{(N,S,\alpha,\sigma)}(x,x')&=&
-\frac{1}{\alpha (2\pi)^{(N+1)/2}}
\int_{-\infty}^{\infty}d\omega\
e^{-i\omega(t-t')}
\sum_{m=-\infty}^{\infty}
e^{i(m+\sigma)\left(\varphi-\varphi'\right)/\alpha}
\nonumber
\\
&&\times\int_{0}^{\infty}d\kappa\ \kappa
J_{|m+\sigma +\omega S|/\alpha}(\kappa\rho)
J_{|m+\sigma +\omega S|/\alpha}(\kappa\rho ')
\nonumber
\frac{\left[\kappa ^{2}+M^{2}-\omega^{2}\right]^{(N-5)/4}}
{\left[({\bf z}-{\bf z'})^{2}\right]^{(N-5)/4}}
\\
&&\times
K_{(N-5)/2}\left(\left\{\left(\kappa^{2}+M^{2}-\omega ^{2}\right)
({\bf z}-{\bf z'})^{2}\right\}^{1/2}\right),
\nonumber
\en
with $K_{\nu}$ denoting a modified Bessel function
of the second kind. In the case of a spinless cone,
\bn
G_{{\cal F}}^{(N,0,\alpha,\sigma)}(x,x')&=&
-\frac{i}{\alpha (2\pi)^{N/2}}
\sum_{m=-\infty}^{\infty}
e^{i(m+\sigma)\left(\varphi-\varphi'\right)/\alpha}
\nonumber\\
&&\times\int_{0}^{\infty}d\kappa\ \kappa
J_{|m+\sigma |/\alpha}(\kappa\rho)
J_{|m+\sigma |/\alpha}(\kappa\rho ')
\frac{\left[{\kappa ^{2}+M^{2}}\right]^{(N-4)/4}}
{\left[({\bf z}-{\bf z'})^{2}-(t-t')^{2}\right]^{(N-4)/4}}
\nonumber
\\
&&\times K_{(N-4)/2}\left(\left\{\left(\kappa^{2}+M^{2}\right)
\left[({\bf z}-{\bf z'})^{2}-(t-t')^{2}\right]
\right\}^{1/2}\right).
\label{pro6}
\en
$G_{{\cal F}}^{(4,0,\alpha,0)}(x,x')$
obtained from (\ref{pro6})
agrees with the Wightman function in
\cite{smi90,shi92}.
So do $G_{{\cal F}}^{(3,0,\alpha,0)}(x,x')$
and $G_{{\cal F}}^{(3,0,\alpha,1/2)}(x,x')$
agree with the Wightman functions in \cite{sou92}.

\section{The Dirac propagator}
Now a spinless cone
will be consider in order to avoid
subtleties concerned with unitarity,
for the spin 1/2 field.
For the same reason $\sigma$ is not
identified with a flux parameter
(see, for example \cite{ger89a,ger89b,ger90}).

To study higher spins on the cone one needs to
use the $N$-bein formalism \cite{wei72,dav82}.
The metric tensor in (\ref {le}) can be
generated by the $N$-bein
\be
e_{a}{} ^{\mu}=
\left(\begin{array}{cccccc}
1&0&0&0&\cdots&0\\
0&\cos(\varphi /\alpha)&-\rho ^{-1}\sin(\varphi /\alpha)&0&\cdots&0\\
0&\sin(\varphi /\alpha)&\rho^{-1}\cos(\varphi /\alpha)&0&\cdots&0\\
0&0&0&1&\cdots&0\\
\vdots&\vdots&\vdots&\vdots&\ddots&\vdots\\
0&0&0&0&\cdots&1
\end{array}
\right),
\label{tet}
\ee
which is continuous on the ray $\varphi=0\sim 2\pi\alpha$.
For $N=3$ $e_{a}{} ^{\mu}$ is the dreibein in \cite{ger89a}
expressed in terms of the coordinate system
$\left(t,\rho ,\varphi\right)$.
For the spin 1/2 field, the spin connection associated with
$e_{a}{} ^{\mu}$ is
\bn
\Gamma _{\mu}&=&\frac{1}{4}\gamma ^{a}\gamma ^{b}
e_{a}{} ^{\nu}e_{b{}\nu{};{}\mu}
\label{sc1}\\
&=&i\delta _{\mu}{} ^{2}(\alpha ^{-1}-1){\cal J},
\nonumber
\en
where $\gamma ^{a}$ are the $\gamma$ matrices and
${\cal J}:=i\gamma ^{1}\gamma ^{2}/2$.
When the conical singularity is absent, i.e. $\alpha=1$,
$\Gamma_{\mu}$ vanishes and
$e_{a}{} ^{\mu}$ becomes the usual $N$-bein
associated with the Minkowski coordinate system.

Likewise the vector potential in the
A-B set up, the spin connection,
can be gauged way by rotating
$e_{a}{} ^{\mu}$ by an angle $[\alpha ^{-1}-1]\varphi$,
\bd
e_{a}{} ^{\mu}\rightarrow \bar{e}_{a}{} ^{\mu}=
\Lambda _{a}{}^{b} ([\alpha ^{-1}-1]\varphi)
e_{b}{} ^{\mu},
\ed
where
$\Lambda _{a}{}^{b} (\theta)$ is the usual rotation matrix.
The new $N$-bein $\bar{e}_{a}{} ^{\mu}$
is given by (\ref{tet}) setting $\alpha=1$
and the new spin connection $\bar{\Gamma}_{\mu}$ can be
obtained from (\ref{sc1}).
Alternatively one can use the transformation law
$\bar{\Gamma} _{\mu}=
U(\theta)\Gamma_{\mu}U^{-1}(\theta)
-U^{-1}(\theta)
\partial_{\mu}U(\theta)$,
with
\footnote{$U(\theta)$ is the operator corresponding to
a rotation by an angle $\theta$.}
$U(\theta)
=\exp\{i\theta {\cal J}\}$
for $\theta=(\alpha ^{-1}-1)\varphi$,
finding that $\bar{\Gamma}_{\mu}$ vanishes everywhere
except on the ray $\varphi=0\sim 2\pi\alpha$, where
$\bar{e}_{a}{} ^{\mu}$
is discontinuous (for $\alpha\neq 1$).
When $\alpha=1$,
$\bar{e}_{a}{} ^{\mu}\equiv e_{a}{} ^{\mu}$,
and  $\Lambda _{a}{}^{b} ([\alpha ^{-1}-1]\varphi)$
and $U([\alpha^{-1}-1]\varphi)$
become the unity transformations.

The spin 1/2 Feynman propagator on the cone satisfies \cite{dav82}
\bd
\left(i\not\!\nabla _{x}-M\right)S_{{\cal F}}(x,x')=
\frac{1}{\rho}\delta\left(x-x'\right),
\ed
with
$\not\!\nabla:=\gamma^{a} e_{a}{} ^{\mu}
(\partial_{\mu} + \Gamma_{\mu})$,
and the boundary condition (\ref{bc1}).
Thus in terms of $\bar{e}_{a}{} ^{\mu}$
the new Feynman propagator,
$\bar{S}_{{\cal F}}(x,x')=
U([\alpha^{-1}-1]\varphi)
S_{{\cal F}}(x,x')
U^{-1}([\alpha^{-1}-1]\varphi ')$,
satisfies the usual equation in Minkowski spacetime
with the boundary condition
\be
\bar{S}_{{\cal F}}(\varphi+2\pi\alpha)=
\exp\{i2\pi\sigma\}U({\cal D})\bar{S}_{{\cal F}}(\varphi).
\label{bc2}
\ee
It follows that
\be
\bar{S}_{{\cal F}}(x,x')=
\left(i\not\!\partial _{x}+M\right)\bar{G}(x,x')
\label{S}
\ee
with
$\not\!\partial:=\gamma^{a} \bar{e}_{a}{} ^{\mu}
\partial_{\mu} $
and the bispinor $\bar{G}(x,x')$ satisfying
(\ref{fe0}) and (\ref{bc2}).

To obtain $\bar{G}(x,x')$ one can proceed as
in the scalar case.
The eigenfunctions of
$\Box_{x}+M^{2}$
satisfying (\ref{bc2}) are given by
\bn
\Psi_{\omega,\kappa,m,{\bf k}}(x)&=&
\left\{\frac{1}{2}\left[e^{i(1-\alpha^{-1})\varphi/2}
\psi^{+}_{\omega,\kappa,m,{\bf k}}(x)+
e^{i(\alpha^{-1}-1)\varphi/2}
\psi^{-}_{\omega,\kappa,m,{\bf k}}(x)\right]\right.
\nonumber\\
&&\quad +\left[e^{i(1-\alpha^{-1})\varphi/2}
\psi^{+}_{\omega,\kappa,m,{\bf k}}(x)-
e^{i(\alpha^{-1}-1)\varphi/2}
\psi^{-}_{\omega,\kappa,m,{\bf k}}(x)\right]
{\cal J} \biggr\}
U([\alpha^{-1}-1]\varphi),
\nonumber
\en
where
$\psi^{+}$
and
$\psi^{-}$
are c-number
eigenfunctions of
$\Box_{x}+M^{2}$
twisted by angles
$2\pi\sigma+{\cal D}/2$
and
$2\pi\sigma-{\cal D}/2$
respectively.
Then one has that
$\bar{G}(x,x')$
is given by (\ref{pro1})
replacing $\psi $ and $\psi^{\ast}$
by $\Psi$ and $\Psi^{\dagger}$,
\bn
&&\bar{G}^{(N,0,\alpha,\sigma)}(x,x')=
U([\alpha^{-1}-1](\varphi-\varphi '))
\label{M}\\
&&\times\left\{\frac{1}{2}\left[e^{i(1-\alpha^{-1})(\varphi-\varphi ')/2}
G_{{\cal F}}^{(N,0,\alpha,\sigma+{\cal D}/4\pi)}(x,x')+
e^{i(\alpha^{-1}-1)(\varphi-\varphi ')/2}
G_{{\cal F}}^{(N,0,\alpha,\sigma-{\cal D}/4\pi)}(x,x')\right]
\right.\nonumber\\
&&\quad +\left[e^{i(1-\alpha^{-1})(\varphi-\varphi ')/2}
G_{{\cal F}}^{(N,0,\alpha,\sigma+{\cal D}/4\pi)}(x,x')-
e^{i(\alpha^{-1}-1)(\varphi-\varphi ')/2}
G_{{\cal F}}^{(N,0,\alpha,\sigma-{\cal D}/4\pi)}(x,x')\right]
{\cal J} \biggr\}.
\nonumber
\en
Therefore the Dirac propagator
$\bar{S}_{{\cal F}}^{(N,0,\alpha,\sigma)}(x,x')$
is determined by the Klein-Gordon propagator
$G_{{\cal F}}^{(N,0,\alpha,\sigma \pm {\cal D}/4\pi)}(x,x')$
through (\ref {S}).
The second term in
the curly brackets of (\ref{M})
is only due to the conical singularity
and ${\cal D}/4\pi$
plays a role similar to the one played
by the flux parameter in the A-B set up.

To get from (\ref{M}) to the corresponding expressions in terms of
$e_{a}{} ^{\mu}$ and of the polar $N$-bein
$\tilde{e}_{a}{} ^{\mu}=\Lambda _{a}{}^{b} (\varphi/\alpha)
e_{b}{} ^{\mu}$,
one drops the factors
$U([\alpha^{-1}-1](\varphi-\varphi '))$ and
$U(-(\varphi-\varphi '))$ respectively.

Clearly when $\alpha =1$ $({\cal D}=0)$ and
$\sigma =0$, (\ref{M}) collapses into
(\ref{ur}) and (\ref{S}) yields the familiar
Dirac propagator in Minkowski spacetime.

\section{Vacuum expectation values}
Propagators can be used to work out
vacumm expectation values of physical
quantities since the former
are defined as the vacuum expectation
values of products of operators
\cite{dav82,ful89}.
In the following some vacuum expectation
values are determined. It is assumed that
the cone is spinless.

A simple
example is the vacuum fluctuation of a
scalar field
\be
\langle\phi ^{2}(x)\rangle=
i\lim_{x'\rightarrow x}G_{{\cal F}}(x,x'),
\label{vf}
\ee
which obviously diverges.
On a conical background however, because the divergences
have the same nature as
the Minkowski ones, renormalization
can be performed simply by subtracting the
untwisted contribution in Minkowski spacetime
($\alpha=1$, $\sigma =0$)
\footnote{In \cite{mat90a} renormalization is effected
by subtracting the contribution for
$\alpha=1$ and $\sigma =1/2$. In so doing
the twist Casimir effect is missed.
It should be recalled that $\sigma $ can
be thought of as a flux parameter.}.

In order to evaluate (\ref{vf}) one sets
in (\ref{pro3}) $t=t'$, $\rho=\rho'$,
${\bf z}={\bf z}'$ and performs the integration over
$T$ \cite{pru86},
\bn
&&G_{{\cal F}}^{(N,0,\alpha,\sigma)}(\Delta)=
\frac{1}{i\alpha(2\pi^{1/2})^{N}\rho^{N-2}}\times
\nonumber
\\
&&\quad\qquad\sum_{m=-\infty}^{\infty}
e^{i(m+\sigma)\Delta/\alpha}
\left\{\frac{\Gamma\left[(N-2)/2+|m+\sigma|/\alpha\right]
\Gamma\left[(3-N)/2\right]}
{\pi^{1/2}\Gamma\left[(4-N)/2+|m+\sigma|/\alpha\right]}
\nonumber\right.\times
\\
&&\quad\qquad\left. {}_{1}F _{2}\left[(3-N)/2;(4-N)/2-|m+\sigma|/\alpha ,
(4-N)/2+|m+\sigma|/\alpha ;(M\rho)^{2}\right]\nonumber\right.
\\
&&\quad\qquad\left.+\;2^{-2|m+\sigma|/\alpha}
(M\rho)^{N-2(1-|m+\sigma|/\alpha)}
\frac{\Gamma\left[(2-N)/2-|m+\sigma|/\alpha\right]}
{\Gamma\left[1+|m+\sigma|/\alpha\right]}\nonumber\right.\times
\\
&&\quad\qquad{}_{1}F_{2}\left[1/2+|m+\sigma|/\alpha ;
|m+\sigma|/\alpha+N/2 ,1+2|m+\sigma|/\alpha ;(M\rho)^{2}\right]
\biggr\},
\label{del}
\en
where $\Delta:=\varphi-\varphi'$ and
${}_{1}F_{2}\left[a ;b,c;z\right]$
denotes the generalized hypergeometric function,
which converges for all values of $z$ \cite{sla66}.

By taking
$M\rho\rightarrow 0$
in (\ref{del}),
\be
D_{{\cal F}}^{(N,0,\alpha,\sigma)}(\Delta)=
\frac{\Gamma\left[(3-N)/2\right]}
{i\alpha(2\pi^{1/2})^{N}\pi^{1/2}\rho^{N-2}}
\sum_{m=-\infty}^{\infty}
\frac{\Gamma\left[(N-2)/2+|m+\sigma|/\alpha\right]}
{\Gamma\left[(4-N)/2+|m+\sigma|/\alpha\right]}
e^{i(m+\sigma)\Delta/\alpha},
\label{mdel}
\ee
which is the massless propagator. Therefore in
the regime $M\rho<<1$, the vacuum fluctuations
behave approximately as the massless ones.
Later it will be seen that this is also
the case for the energy momentum tensor.
This fact is also mentioned in
\cite{shi92} for $N=4$.

A superficial inspection of (\ref{del})
and (\ref{mdel}) reveals the
usual divergence for $N=2$
when $\sigma$ is an integer
and another for $N=3$.
In (\ref{mdel}) for example,
the $N=2$ divergence is in the term
corresponding to $m=-\sigma$
which is an infinite constant.
For $N=3$,
$\Gamma(0)\delta(\Delta)$
arises in (\ref{mdel}) which
can be regularized by
considering that
$\Gamma[a+z]/\Gamma[b+z]=
B[z+a,b-a]/\Gamma[b-a]$.
By using the usual integral
representation of the beta
function
$B[z,w]$,
the sum in
(\ref{mdel}) can be
evaluated giving a integral
representation for
$D_{{\cal F}}^{(N,0,\alpha,\sigma)}(\Delta)$
which is finite for $N=3$, diverging however
for $N=4$. This representation
will not be given here
\footnote{Alternatively
another representation
for $G_{{\cal F}}$ can be used.
For example representation (\ref{pro6}) for
$G_{{\cal F}}^{(3,0,\alpha,0)}(x,x')$
and
$G_{{\cal F}}^{(3,0,\alpha,1/2)}(x,x')$
is used in \cite{sou92}.}.

{}From (\ref{vf}) and (\ref{mdel}), one sees that
$\langle\phi ^{2}(x)\rangle^{(N,0,\alpha,\sigma)}$
behave as
$1/\rho^{N-2}$.
$\langle\phi ^{2}(x)\rangle^{(4,0,\alpha,\sigma)}$
will now be determined, exhibiting the dependence on
$\sigma$ and $\alpha$.
Taking $|\sigma|\leq 1$ is enough
to cover all the possibilities in (\ref{bc1}).
Setting $N=4$ in (\ref{mdel}),
\bn
D_{{\cal F}}^{(4,0,\alpha,\sigma)}(\Delta)&=&
\frac{i}{2(2\pi\alpha\rho)^{2}}
\sum_{m=-\infty}^{\infty}
|m+\sigma|e^{i(m+\sigma)\Delta/\alpha}
\nonumber
\\
&=&\frac{i}{2(2\pi\alpha\rho)^{2}}
e^{i\sigma\Delta/\alpha}
\left[|\sigma|+
2\sum_{m=1}^{\infty}m\cos \frac{m\Delta}{\alpha}
+2i\sigma\sum_{m=1}^{\infty}\sin \frac{m\Delta}{\alpha}\right]
\nonumber
\\
&=&\frac{i}{2(2\pi\alpha\rho)^{2}}
e^{i\sigma\Delta/\alpha}
\left[|\sigma|-\frac{1}{2}\csc ^{2}\frac{\Delta}{2\alpha}
+i\sigma\cot\frac{ \Delta}{2\alpha}\right],
\label{mdel4s}
\en
which holds for
$|\sigma|\leq 1$.
$D_{{\cal F}}^{(4,0,\alpha,0)}(\Delta)$
and
$D_{{\cal F}}^{(4,0,\alpha,1/2)}(\Delta)$
reproduce the results in \cite{hel86}
and in \cite{mat90a} respectively.
Expanding (\ref{mdel4s}) in powers of $\Delta$,
one obtains
from (\ref{vf}) after renormalization
\be
\langle\phi ^{2}(x)\rangle^{(4,0,\alpha,\sigma)}=
-\frac{1}{48\pi^{2}\rho^{2}}
\left\{\alpha^{-2}\left[6|\sigma|(1-|\sigma|)-1\right]+1\right\}.
\label{vfr}
\ee
$\langle\phi ^{2}(x)\rangle^{(4,0,\alpha,0)}$
and
$\langle\phi ^{2}(x)\rangle^{(4,0,\alpha,1/2)}$
agree with the results in \cite{smi90}
and
$\langle\phi ^{2}(x)\rangle^{(4,0,1,1/2)}$
agrees with the result in \cite{for80}.
The absolute value of $\sigma$ indicates
that the direction of the twist
(magnetic flux
\footnote{$\phi(x)$ is hermitian. The expectation values
of a charged scalar field are twice the
ones of $\phi(x)$ \cite{dow87a}.})
is irrelevant. The
divergence at $\rho=0$
is due to the conical singularity
and/or the twist \cite{smi90}.

In order to obtain mass corrections to (\ref{vfr})
one has to keep terms of order $(M\rho)^{2}$ in (\ref{del}).
Only the untwisted case will be considered
(i.e., $\sigma=0$). For $N=4$ divergences
arise in the $(M\rho)^{2}$ term. A convenient procedure
to deal with such divergences is dimensional
regularization since (\ref{del}) is expressed in
arbitrary dimensions. Setting $N=4-\epsilon$,
one expands the coefficient of $(M\rho)^{2}$
in powers of $\epsilon$, isolating the singularities
$1/\epsilon$ which cancel out. Then taking
$\epsilon\rightarrow 0$,
\bn
G_{{\cal F}}^{(4,0,\alpha,0)}(\Delta)&=&
-\frac{i}{4(2\pi\alpha\rho)^{2}}
\csc ^{2}\frac{\Delta}{2\alpha}
\nonumber
\\
&&-\frac{iM^{2}}{8\alpha\pi^{2}}
\left[\log M\rho+\alpha\log \sin\frac{\Delta}{2\alpha}
+\gamma-\frac{1}{2}+(\alpha-1)\log 2\right],
\nonumber
\en
where $\gamma$ is the Euler constant and for which
$G_{{\cal F}}^{(4,0,1,0)}(\Delta)$ reproduces
the corresponding expansion for the
massive Klein-Gordon propagator
in Minkowski spacetime.
Then (\ref{vf}) gives
\bn
\langle\phi ^{2}(x)\rangle^{(4,0,\alpha,0)}&=&
\frac{1}{48\pi^{2}\rho^{2}}
\left(\alpha^{-2}-1\right)
\label{mvfr}
\\
&&+\frac{M^{2}}{8\pi^{2}}
\left[\left(\alpha^{-1}-1\right)\left(\log \frac{M\rho}{2}
+\gamma-\frac{1}{2}\right)-\log\alpha\right],
\nonumber
\en
reducing to (\ref{vfr}) when $M\rho\rightarrow 0$.
The massless term in
(\ref{mvfr}) diverges positively at the conical singularity,
whereas the massive term diverges negatively.

A similar procedure can be used to obtain
the vacuum expectation value of the energy
momentum tensor. One can show that in general
\be
\langle T^{\mu}{}_{\nu}(x)\rangle=
i\lim_{x'\rightarrow x}{\rm D}^{\mu}{}_{\nu}(x,x')
G_{{\cal F}}(x,x'),
\label{st}
\ee
where
\bd
{\rm D}^{\mu}{}_{\nu}(x,x'):=
(1-2\xi)\nabla^{\mu}\nabla_{\nu '}
+(2\xi-1/2)\delta^{\mu}{}_{\nu}
\nabla_{\sigma}\nabla^{\sigma '}
-2\xi\nabla^{\mu}\nabla_{\nu }
+2M^{2}\delta^{\mu}{}_{\nu}
(1/4-\xi),
\ed
corresponds to the classical energy
momentum tensor of a scalar field
coupled to the curvature scalar
with coupling parameter
$\xi$ \cite{dav82,ful89}.
The prime in $\nabla_{\nu '}$ indicates
that the covariant derivative is
taken with respect to $x^{\nu'}$.
Though the coupling with the
curvature scalar does not affect
the field equation (\ref{fe0}),
it does affect
$\langle T^{\mu}{}_{\nu}(x)\rangle$
through $\xi$,
as is remarked in \cite{smi90}.

One can express (\ref{st})
for a conical background and with
$M\rho<<1$ in terms of
$D_{{\cal F}}(\Delta)$ and
$\partial_{\varphi}^{2}D_{{\cal F}}(\Delta)$
only. Using the relations
\bn
&&\lim_{x'\rightarrow x}
\partial_{\varphi}\partial_{\varphi '}
G_{{\cal F}}(x,x')=
-\lim_{\Delta\rightarrow 0}\partial^{2}_{\varphi}
G_{{\cal F}}(\Delta)
\label{rel1}
\\
&&\lim_{x'\rightarrow x}
\partial_{z_{i}}\partial_{z_{i}'}
G_{{\cal F}}(x,x')=
-\lim_{x'\rightarrow x}
\partial_{z_{i}}^{2}
G_{{\cal F}}(x,x')=
-\lim_{x'\rightarrow x}\partial_{t}
\partial_{t'}G_{{\cal F}}(x,x')=
\lim_{x'\rightarrow x}\partial_{t}^{2}
G_{{\cal F}}(x,x'),
\nonumber
\en
which can be derived from (\ref{pro3}), and
\bn
&&\lim_{x'\rightarrow x}\partial_{\rho}
G_{{\cal F}}(x,x')=
\frac{2-N}{2\rho}
\lim_{\Delta\rightarrow 0}
D_{{\cal F}}(\Delta)
\label{rel2}
\\
&&\lim_{x'\rightarrow x}\partial_{t}
\partial_{t'}G_{{\cal F}}(x,x')=
\frac{1}{(1-N)\rho^{2}}
\lim_{\Delta\rightarrow 0}
\left[\left(\frac{N-2}{2}\right)^{2}+
\partial_{\varphi}^{2}\right]D_{{\cal F}}(\Delta)
\nonumber
\\
&&\lim_{x'\rightarrow x}\partial_{\rho}
\partial_{\rho'}G_{{\cal F}}(x,x')=
\frac{1}{(N-1)\rho^{2}}
\lim_{\Delta\rightarrow 0}
\left[N\left(\frac{N-2}{2}\right)^{2}+
\partial_{\varphi}^{2}\right]D_{{\cal F}}(\Delta)
\nonumber
\\
&&\lim_{x'\rightarrow x}
\partial_{\rho}^{2}G_{{\cal F}}(x,x')=
\frac{1}{(N-1)\rho^{2}}
\lim_{\Delta\rightarrow 0}
\left\{\frac{(N-2)}{2}\left[\frac{N(N-2)}{2}+1\right]
-\partial_{\varphi}^{2}\right\}D_{{\cal F}}(\Delta)
\nonumber
\en
which can be obtained by differentiating
(\ref{pro3}), and by evaluating the integration
over $T$
taking
$M\rho\rightarrow 0$
it follows that
\bn
&&\langle T^{\mu}{}_{\nu}(x)\rangle
^{(N,0,\alpha,\sigma)}=
\frac{i}{\rho^{2}}\left[\frac{1}{1-N}
\lim_{\Delta\rightarrow 0}
\partial_{\varphi}^{2}D_{{\cal F}}
^{(N,0,\alpha,\sigma)}(\Delta)\,
{\rm diag}(1,1,1-N,1,\ldots,1)\right.
\nonumber
\\
&&+(N-2)(\xi -\xi_{N})
\lim_{\Delta\rightarrow 0}
D_{{\cal F}}
^{(N,0,\alpha,\sigma)}(\Delta)\,
{\rm diag}(2-N,1,1-N,2-N,\ldots,2-N)\biggr],
\label{str}
\en
with $\xi_{N}:=(N-2)/4(N-1)$.
Thus
$\langle T^{\mu}{}_{\nu}(x)\rangle$ behaves as
$1/\rho^{N}$,
is covariantly conserved and traceless when $\xi=\xi_{N}$.
For $N=4$ and $N=3$ respectively, the results in \cite{hel86} and
\cite{sou92} are in agreement with (\ref{rel1}), (\ref{rel2})
and (\ref{str}).

To demonstrate the dependence
on $\sigma$ and $\alpha$,
$\langle T^{\mu}{}_{\nu}(x)\rangle
^{(4,0,\alpha,\sigma)}$
is given by,
\bn
&&\langle T^{\mu}{}_{\nu}(x)\rangle
^{(4,0,\alpha,\sigma)}=
\frac{1}{1440\pi^{2}\rho^{4}}
\Bigl\{\left\{\alpha^{-4}\left[30\sigma^{2}
(1-|\sigma|)^{2}-1\right]+1\right\}
{\rm diag}(1,1,-3,1)
\nonumber
\\
&&\quad\qquad\qquad\qquad+10(6\xi-1)
\left\{\alpha^{-2}
\left[6|\sigma|(1-|\sigma|)-1\right]+1\right\}
{\rm diag}(2,-1,3,2)\Bigr\},
\label{st4}
\en
where (\ref{mdel4s}) has been used.
This result is in agreement with
\cite{fro87} and \cite{dow87b}.
$\langle T^{\mu}{}_{\nu}(x)\rangle
^{(4,0,\alpha,0)}$ has been evaluated
in the literature by different methods
\cite{hel86,lin87,smi90}, and
$\langle T^{\mu}{}_{\nu}(x)\rangle
^{(4,0,1,1/2)}$ and
$\langle T^{\mu}{}_{\nu}(x)\rangle
^{(4,0,\alpha,1/2)}$
have been determined in
\cite{for80} and \cite{smi90}
respectively.
Clearly mass corrections to (\ref{st4})
can be obtained in the same way as in (\ref{mvfr}).
They will not be given here.

For the spin 1/2 field where
the propagators are defined in terms
of anticommuting fields, it can be shown that
the vacuum expectation value corresponding
to the classical energy momentum tensor
\cite{dav82} is
\bd
\langle T_{\mu}{}_{\nu}(x)\rangle=
\frac{1}{4}\lim_{x'\rightarrow x}{\rm Tr}
\left[\bar{\gamma}_{\mu}(\partial_{\nu}-\partial_{\nu'})
+\bar{\gamma}_{\nu}(\partial_{\mu}-\partial_{\mu'})\right]
\bar{S}_{{\cal F}}(x,x'),
\ed
where
$\bar{\gamma}_{\mu}=
\bar{e}_{a}{} _{\mu}\gamma^{a}$.
For $M\rho<<1$
a general expression such as (\ref{str})
can also be obtained
from (\ref{rel1}) and (\ref{rel2}). It
will not be given here.
Only the energy density
$\langle T_{0}{}_{0}(x)\rangle
^{(4,0,\alpha,\sigma)}$
will be given.
Using (\ref{S}), (\ref{M}) and (\ref{pro3}),
\bd
\langle T_{0}{}_{0}(x)\rangle
^{(N,0,\alpha,\sigma)}=
i\lim_{x'\rightarrow x}{\rm Tr}
\left\{\partial_{t}^{2}
\bar{G}^{(N,0,\alpha,\sigma)}(x,x')\right\}.
\nonumber
\ed
Observing the properties of the trace
of the $\gamma$ matrices, and also
(\ref{rel1}), (\ref{rel2}) and
(\ref{mdel4s}),
\bn
&&\langle T_{0}{}_{0}(x)\rangle
^{(4,0,\alpha,\sigma)}=
-\frac{1}{12\pi^{2}\rho^{4}}
\left\{\frac{\alpha^{-4}}{2}
\left[\sigma_{+}^{2}
(1-|\sigma_{+}|)^{2}+
\sigma_{-}^{2}(1-|\sigma_{-}|)^{2}
-\frac{1}{15}\right]\right.
\nonumber
\\
&&\qquad\qquad\qquad
\qquad\qquad\qquad
\left.\ +\alpha^{-2}
\left[|\sigma_{+}|
(1-|\sigma_{+}|)+
|\sigma_{-}|(1-|\sigma_{-}|)
-\frac{1}{3}\right]
+\frac{11}{30}\right\}
\nonumber
\en
where
$\sigma_{\pm}:=\sigma\pm{\cal D}/4\pi$
and which holds for
$|\sigma_{\pm}|\leq 1$.
Two cases can be of particular
interest,
the Casimir effect due to the
conical singularity only
\bd
\langle T_{0}{}_{0}(x)\rangle
^{(4,0,\alpha,0)}=
-\frac{1}{2880\pi^{2}\rho^{4}}
(\alpha^{-2}-1)(7\alpha^{-2}+17)
\ed
in agreement with \cite{fro87},
and
\bd
\langle T_{0}{}_{0}(x)\rangle
^{(4,0,1,\sigma)}=
\frac{1}{12\pi^{2}\rho^{4}}
|\sigma|(\sigma^{2}-1)
(2-|\sigma|),
\ed
which is the Casimir effect caused
by the twist only.

\section{Conclusion and remarks}
Summarizing,
representations of the Feynman propagator
$G_{{\cal F}}^{(N,S,\alpha,\sigma)}(x,x')$
of a massive scalar field ($M$) twisted by
an angle $2\pi\sigma$
have been determined
in a background of a $N$ dimensional spinning cone ($S$)
of deficit angle
${\cal D}= 2\pi \left(1-\alpha \right)$.
These representations are a generalization
of those in the literature. It has been
shown that the Dirac propagator
$S_{{\cal F}}^{(N,0,\alpha,\sigma)}(x,x')$
can be obtained from
$G_{{\cal F}}^{(N,0,\alpha,\sigma\pm{\cal D}/4\pi)}(x,x')$.
The calculation of
$S_{{\cal F}}^{(N,0,\alpha,\sigma)}(x,x')$
has been performed
in a gauge in which the spin connection
vanishes everywhere except on a ray.
Various Aharonov-Bohm like features as
this one have been discussed.

In the coincidence limit
$t=t'$, $\rho=\rho'$,
${\bf z}={\bf z}'$,
$G_{{\cal F}}^{(N,0,\alpha,\sigma)}(\Delta)$
has been expressed in terms
of series in powers of $M\rho$
reducing, when
$M\rho<<1$,
to the massless expression
$D_{{\cal F}}^{(N,0,\alpha,\sigma)}(\Delta)$,
which behaves as
$1/\rho^{N-2}$.
Using this series representation,
vacuum expectation values
have been evaluated.
$\langle\phi ^{2}(x)\rangle^{(4,0,\alpha,\sigma)}$
has been obtained
and the mass correction in
$\langle\phi ^{2}(x)\rangle^{(4,0,\alpha,0)}$
has shown to have a logarithmic divergence
at the conical singularity.

The coincidence limit of the derivatives of
$G_{{\cal F}}^{(N,0,\alpha,\sigma)}(\Delta)$
in the formulas for
the energy momentum tensors
$\langle T^{\mu}{}_{\nu}(x)\rangle
^{(N,0,\alpha,\sigma)}$
have been
written in terms of
$D_{{\cal F}}(\Delta)$ and
$\partial_{\varphi}^{2}D_{{\cal F}}(\Delta)$
when
$M\rho<<1$.
As a consequence,
a general expression for
$\langle T^{\mu}{}_{\nu}(x)\rangle
^{(N,0,\alpha,\sigma)}$
of a scalar field has been
given. It behaves as $1/\rho^{N}$, satisfies
the covariant conservation law
and is traceless for the conformal
coupling with the curvature scalar.
It has been remarked that a similar general
expression can also be obtained for spinors.
$\langle T^{\mu}{}_{\nu}(x)\rangle
^{(4,0,\alpha,\sigma)}$ of a scalar
field and
$\langle T_{0}{}_{0}(x)\rangle
^{(4,0,\alpha,\sigma)}$ of a
spinor field have been evaluated,
showing the dependence on the parameters
$\sigma$ and $\alpha$.

Comparing the expressions in section 3
with those in section 4, one sees
a loose correspondence,
$\omega\times S \leftrightarrow
{\cal D}/4\pi = \mu\times1/2$.
In the context of first quantization
in three dimensions ($N=3$), it has
been found \cite{ger89a} that
$\omega \leftrightarrow\mu$
and
$S\leftrightarrow s$,
where $s$ is the spin of
the particle.
If these correspondences are
carried over to the context of
second quantization, one concludes
that the factor $1/2$ above
refers to the spin of the field.
In the light of this remark and
that of \cite{dow87b} one is led
to conjecture that formula
(\ref{M}) also holds for arbitrary
spins, replacing
$G_{{\cal F}}^{(N,0,\alpha,\sigma\pm{\cal D}/4\pi)}(x,x')$
by
$G_{{\cal F}}^{(N,0,\alpha,\sigma\pm{\cal D}s /2\pi)}(x,x')$
and with
${\cal J}$
being the spin $s$ generator of a rotation.
The $\gamma$ matrices would be replaced by
the ones mentioned in \cite{dow87b}.
Thus the Feynman propagator of
a spin $s$ field twisted by
an angle $2\pi\sigma$ would be obtained
from the scalar propagator twisted
by angles $2\pi\sigma\pm{\cal D}s$.
But this deserves further investigation.

Another extension of this work is to
consider
$S_{{\cal F}}^{(N,S,\alpha,\sigma)}(x,x')$
when $S\neq 0$ and with flux parameter $\sigma$.
In doing so one must face the
problems mentioned earlier
concerning unitarity.
Also worth investigating is
the behaviour of the vacuum
expectation values when $S\neq 0$.
Using the asymtotic limit
of
${}_{1}F_{2}\left[a ;b,c;z\right]$
for large $z$, they
can in principle be evaluated also when
$M\rho>>1$.
The numerical analysis
of the vacuum expectation
values of a scalar field in
$N=3$
\cite{sou92,mak93}
can be extended to spinors
through (\ref{M}).

\vspace{10 mm}
{\bf Note}. After this work was completed
references \cite{emi94,lin94} appeared that
also consider massive twisted fields in a $N$
dimensional conical background using different
methods.

\vspace{5 mm}
{\bf Acknowledgements}.
The author is grateful to Michael Green
for advice, to Paul Wai, Jos\'{e} Figueroa
and Geoffrey Sewell for
clarifying discussions, and to
Bobby Acharya for reading the manuscript.

\end{document}